\documentclass{nature}

\usepackage{graphicx}
\usepackage[font=small]{caption}
\usepackage{xcolor}
\usepackage{amsmath} 
\usepackage{booktabs}
\usepackage{float} 

\usepackage{multirow}%
\usepackage{amsmath,amssymb,amsfonts}%
\usepackage{amsthm}%
\usepackage{mathrsfs}%
\usepackage[title]{appendix}%
\usepackage{textcomp}%
\usepackage{manyfoot}%
\usepackage{booktabs}%
\usepackage{algorithm}%
\usepackage{algorithmicx}%
\usepackage{algpseudocode}%
\usepackage{listings}%
\usepackage{braket}
\usepackage{url}
\usepackage{lineno}
\usepackage{amsmath}
\usepackage{tabularx}
\usepackage{makecell}
\usepackage{array}
\usepackage{graphicx}

 \title{Observation of light-induced topological state in graphene}

\author
{Sotirios Fragkos$^{1,*}$, Ann Chantal Goutier$^{2,3}$, Umberto Dellasette$^{1}$, Dominique Descamps$^{1}$, Stéphane Petit$^{1}$, Wiebke Bennecke$^{4}$, Philip Hofmann$^{5}$, Marcel Reutzel$^{6,7}$, Stefan Mathias$^{4,8}$, Yann Mairesse$^{1}$, Michael Schüler$^{2,3,\dag}$, and Samuel Beaulieu$^{1,\ddag}$}

\usepackage[colorlinks,linkcolor={blue!80!black},citecolor={blue!50!black},urlcolor={blue!80!black}]{hyperref}

\begin{document}

\maketitle

\begin{affiliations}
 \item Universit\'e de Bordeaux - CNRS - CEA, CELIA, UMR5107, F33405 Talence, France.
  \item PSI Center for Scientific Computing, Theory and Data, 5232 Villigen PSI, Switzerland.
 \item Department of Physics, University of Fribourg, CH-1700 Fribourg, Switzerland.
\item   I. Physikalisches Institut, Georg-August-Universität Göttingen, Friedrich-Hund-Platz 1, D-37077 Göttingen, Germany. 

\item Department of Physics and Astronomy, Aarhus University, 8000 Aarhus C, Denmark.

\item Fachbereich Physik, Philipps-Universität Marburg, 35032 Marburg, Germany.
\item mar.quest | Marburg Center for Quantum Materials
and Sustainable Technologies, 35032 Marburg, Germany
\item International Center for Advanced Studies of Energy Conversion (ICASEC), University of Göttingen, Göttingen, Germany \\

  $^{*}$sotirios.fragkos@u-bordeaux.fr,
  $^{\dag}$michael.schueler@psi.ch, 
  $^{\ddag}$samuel.beaulieu@u-bordeaux.fr
  
\end{affiliations}

\newpage

\begin{abstract}
Driving quantum materials with light can reshape not only their electronic band structures but also the geometry of their electronic wavefunctions. In graphene, circularly polarized light is predicted to open gaps at the Dirac points and induce a pseudospin texture and Berry curvature characteristic of the Haldane model, realizing a Floquet Chern insulator~\cite{Oka09, Oka19, Lindner2011-oe, Haldane1988}. Transport and all-optical measurements have provided evidence for a light-induced anomalous Hall response~\cite{McIver20, Lesko26}, but cannot resolve the underlying electronic states in both energy and momentum space. Time- and angle-resolved photoemission spectroscopy has revealed Floquet-Bloch sidebands~\cite{merboldt24, choi24} and hybridization gaps at crossings between Floquet bands~\cite{Wang2026-se}. However, these experiments used linearly polarized driving, which preserves time-reversal symmetry and thus does not induce a topological phase. Here, we combine circular-dichroic time-resolved extreme-ultraviolet momentum microscopy~\cite{Fragkos2025} with ab initio theory to investigate graphene driven by circularly polarized light. We observe a pump-induced photoemission circular dichroism near the Dirac point that reverses sign with pump helicity and across the Dirac energy, in agreement with our calculations. These results link the measured dichroic response to the light-induced pseudospin texture and its associated Berry curvature, providing spectroscopic evidence for the Floquet-engineered topological insulating state in graphene.
\end{abstract}

\newpage

The field of Floquet-Bloch engineering~\cite{Torre21, Basov17, Oka19}, which exploits periodic optical driving to tailor a material’s electronic structure and topology, was propelled by Oka and Aoki's pioneering proposal~\cite {Oka09}. They predicted that circularly polarized light can induce a Haldane-like effective Hamiltonian~\cite{Haldane1988} in monolayer graphene, giving rise to a Floquet topological insulator. In this scheme, the periodic drive breaks time-reversal symmetry and opens a gap at the Dirac points. Such Floquet topological states have since been experimentally realized in cold-atom optical lattices and photonic systems~\cite{Jotzu2014, Rechtsman2013}. 

In condensed matter, successful experimental demonstrations of Floquet physics have been largely limited to trivial and topological insulators~\cite{Wang13, Mahmood16, Zhou23, Zhou23_2, Bao2024, Ito23, Bao2024-le, Fragkos25valley}, and, very recently, a transient topological state attributed to a resonantly driven Floquet band inversion was reported in the semiconductor SnTe~\cite{Chassot2026}. In particular, realizing and observing Floquet-Bloch states in graphene has proven challenging~\cite{Aeschlimann21}. Graphene is a semimetal characterized by unusually fast electron-electron and electron-phonon scattering~\cite{Breusing11, Brida2013, Heide2021-mj}, raising the question of whether coherent Floquet dressing can coexist with such rapid scattering and decoherence processes~\cite{Aeschlimann21, Sato2020, Broers2022}. Transport and all-optical measurements reported a light-induced anomalous Hall effect in graphene~\cite{McIver20, Lesko26}. However, because these measurements are integrated over both energy and momentum, they do not allow the topological Floquet–Bloch contribution to be disentangled from effects arising from asymmetric photocarrier distributions~\cite{Sato19}, leaving the microscopic origin of the observed phenomena only partially resolved. Direct spectroscopic evidence of light-induced Floquet topological states in graphene driven using circularly polarized light remains elusive and would require time- and angle-resolved photoemission spectroscopy (trARPES) measurements, which would enable the transient electronic structure to be probed as a function of both energy and momentum.

Very recent trARPES experiments have succeeded in revealing coherent Floquet-Bloch states in graphene under linearly polarized driving by resolving quantum-path interference between Floquet and Volkov sidebands~\cite{merboldt24, choi24}. Floquet-induced hybridization gaps at the crossing between equilibrium and Floquet bands have subsequently been resolved~\cite{Wang2026-se}. However, this avoided-crossing gap is distinct from the Haldane-type mass gap predicted to open at the Dirac points. Indeed, these trARPES experiments~\cite{merboldt24, choi24, Wang2026-se} used linearly polarized driving, which preserves time-reversal symmetry and therefore does not open a Haldane mass at the Dirac points. The predicted Haldane-type gap, together with the associated pseudospin and Berry curvature around the Dirac points, thus remains to be investigated by trARPES.

Merely observing a gap, however, does not prove the existence of a Floquet topological insulator. Indeed, a gap can arise from many mechanisms unrelated to topology~\cite{Balog2010, Hunt13, Gorbachev14} and carries no information on the pseudospin texture and Berry curvature that define the topological phase. The same applies to other band-structure signatures, such as transient band inversions or light-induced Dirac cones~\cite{Chassot2026}: they constrain the spectrum, but the topological character resides in the geometry of the light-dressed wave-functions.
Recent theoretical work has shown that these quantities can be revealed through circular dichroism in trARPES~\cite{Schueler20}. Even when electron-electron and electron-phonon scattering broadens or obscures the Floquet-induced Haldane gap, the sign and momentum dependence of dichroic signals are predicted to retain information about the light-induced pseudospin~\cite{Schueler20}. A related approach was recently demonstrated on the surface of a three-dimensional topological insulator using ultraviolet circular-dichroic trARPES~\cite{Mahmood26}. For graphene, where the Dirac cones lie at the Brillouin-zone boundary, it is necessary to use higher photon energies, which the recent developments of our circularly polarized extreme-ultraviolet (XUV) source make possible~\cite{Fragkos2025}.

Here, we report measurements of pump-helicity-dependent photoemission circular dichroism in monolayer graphene using time- and polarization-resolved XUV momentum microscopy. Comparing the XUV dichroisms obtained with the two circular polarizations of the IR drive, we extract a double- dichroism that reverses sign across the Dirac point, in quantitative agreement with ab initio calculations. This behavior is consistent with the predicted light-induced pseudospin texture and its associated Berry curvature, finally providing direct spectroscopic evidence for a Floquet topological state in graphene.

\subsection{Generation and detection of a light-induced Floquet topological insulator} 
\begin{figure}[H]
\centering
\includegraphics[width=\textwidth]{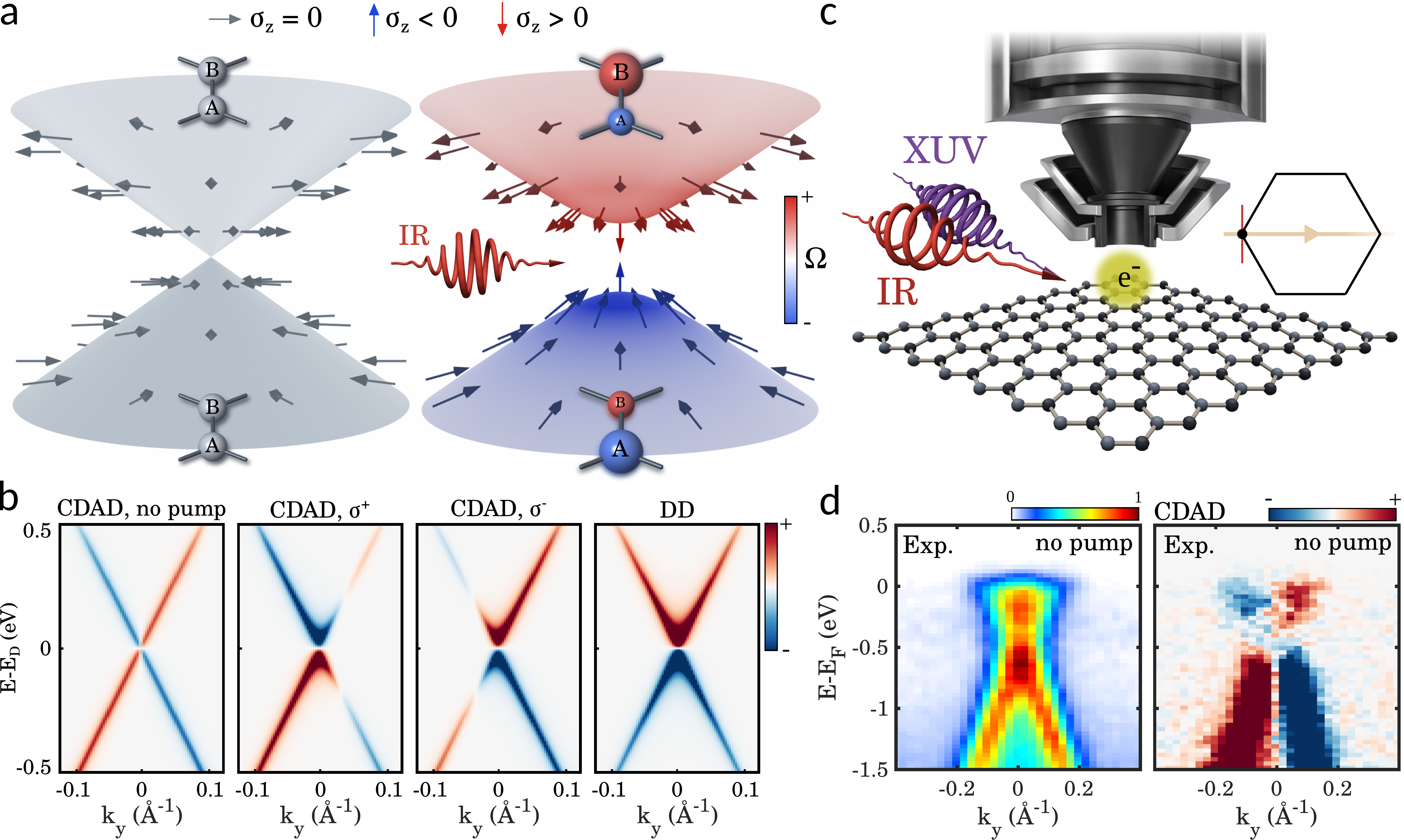}
\caption{\textbf{Light-induced pseudospin, Berry curvature, and static photoemission circular dichroism in graphene} \textbf{(a)} Schematic of the Dirac bands and associated sublattice pseudospin (arrows) in equilibrium (left) and under circularly polarized IR driving (right). In equilibrium, the pseudospin lies in-plane ($\sigma_z=0$) and the A and B sublattice weights are equal. The infrared (IR) driving pulse opens a gap, tilts the pseudospin out-of-plane ($\sigma_z<0$ in blue, $\sigma_z>0$ in red), imbalances the A and B sublattice weights (grey, red/blue spheres) and induces a Berry curvature $\Omega$ (red/blue colorbar) of opposite sign in the two bands. \textbf{(b)} Calculated CDAD for free-standing graphene without pump, with right- ($\sigma^+$) and left- ($\sigma^-$) circularly polarized pump pulses, and the resulting double-helicity dichroism (DD), for a drive strength of 4000~kV/cm. \textbf{(c)} Scheme of the experiment: a polarization-tunable infrared pump (IR, 1.2~eV, 135~fs) and a polarization-tunable XUV probe (21.6~eV) are focused on a graphene/SiC sample in the interaction chamber of a time-of-flight momentum microscope, at an incidence angle of 65$^{\circ}$, with the light incidence plane along $\Gamma$-K. \textbf{(d)} Measured normalized photoemission intensity (left) and CDAD (right) of graphene/SiC without pump, along the momentum direction perpendicular to $\Gamma$-K through the Dirac point.}
\label{Fig1}
\end{figure}

In equilibrium, graphene hosts massless Dirac fermions whose lattice pseudospin lies in-plane and points radially in or out of the Dirac cones (Fig.~\ref{Fig1}a, left), with opposite chirality at K and K$^\prime$. The wavefunction amplitudes on the two sublattices are equal, while their relative phase depends on momentum. In the high-frequency limit, circularly polarized light breaks time-reversal symmetry and generates an effective Haldane mass $m_F$ in graphene~\cite{Oka09,mikami_brillouin-wigner_2016}, opening a gap of $2|m_F|$ at the Dirac points, which scales as $m_F \propto E_0^2/\omega_\mathrm{pump}^3$, where $E_0$ is the electric-field amplitude of the pump pulse and $\omega_\mathrm{pump}$ its angular frequency. The sign of the mass is determined by the pump helicity~\cite{Oka09, Sentef15, Schueler20}.

The out-of-plane component of the pseudospin, $\sigma_z(\mathbf{k}) = P_A(\mathbf{k}) - P_B(\mathbf{k})$, is given by the difference between the probabilities $P_A$ and $P_B$ of finding the electron on the A and B sublattices, respectively, and vanishes in equilibrium (Fig.\ref{Fig1}a, left). Under circularly polarized driving, $\sigma_z$ becomes finite, with opposite signs in the upper and lower bands. Reversing the pump helicity reverses its sign. The resulting pseudospin texture is associated with a finite Berry curvature (Fig.~\ref{Fig1}a, right). This gives rise to a nonzero Chern number $C = (1/2\pi) \int d\mathbf{k}\, \Omega(\mathbf{k}) = 1$, and an associated orbital magnetic moment $M_z(\mathbf{k}) = \mp \tfrac{e}{\hbar} [\varepsilon_+(\mathbf{k}) -\varepsilon_-(\mathbf{k})]\Omega(\mathbf{k})$, where $\varepsilon_{\pm}(\mathbf{k})$ is the upper/lower band, both of which are linked to the light-induced anomalous Hall response~\cite{McIver20}.

The pseudospin texture can be probed using circular dichroism in the angular distribution (CDAD) of photoelectrons emitted from the graphene $\pi$ bands. The photoemission intensity depends on interference between photoemission amplitudes from the two sublattices~\cite{liu_visualizing_2011, Gierz2012,hwang2011}. Under a circularly polarized pump, the leading term of the light-induced intrinsic photoemission XUV circular dichroism is proportional to $\sigma_z(\mathbf{k})$~\cite{Sentef15, Schuler20, Schueler20}. Consequently, CDAD can probe the light-induced sublattice imbalance and, through its relation to the Floquet Hamiltonian, provide information about the associated Berry curvature. As discussed below in connection with Fig.~\ref{Fig4}, the pseudospin extends further away from the Dirac point than the Berry curvature, making the dichroic signal robust against scattering-induced broadening of the Floquet-induced Haldane gap.

The calculated CDAD of free-standing graphene at equilibrium and driven by a circular pump (for both helicities, $\sigma^+$ and $\sigma^-$) are shown in Fig.~\ref{Fig1}b (see Methods). The light-induced contribution is concentrated near the Dirac point, and changes sign with the pump helicity and across the Dirac point. To isolate this contribution, we define the double-helicity dichroism (DD) as $\mathrm{DD}=\mathrm{CDAD}_{\sigma^-}-\mathrm{CDAD}_{\sigma^+}$, following an approach recently used for investigating Floquet-Bloch states on the surface of a three-dimensional topological insulator~\cite{Mahmood26}. This doubly differential observable removes the equilibrium dichroism and other pump-helicity-independent contributions, isolating the helicity-odd response. The resulting DD is expected to be strongest near the Dirac point and to change sign across the Dirac energy (Fig.~\ref{Fig1}b, right).

\subsection{Experimental realization of light-induced topological state.} A schematic of our experimental setup is shown in Fig.~\ref{Fig1}c. In a nutshell, a polarization-tunable infrared pump pulse (1.2~eV, 135~fs, incident fluence 3.5~mJ/cm$^2$) drives monolayer graphene, while a polarization-tunable XUV probe pulse (21.6~eV) ejects photoelectrons. The measurements are performed with the two pulses overlapped in time. The photoelectrons are collected using a time-of-flight momentum microscope, enabling simultaneous measurements across the full accessible energy and in-plane momentum ranges.

We first measure the equilibrium band structure and associated CDAD of unpumped graphene on a SiC substrate. The photoemission intensity along the direction perpendicular to $\Gamma$-K through the Dirac point shows the linear dispersion of the $\pi$ bands (Fig.~\ref{Fig1}d, left). Previous work has shown that static XUV CDAD is sensitive to the lattice pseudospin of graphene~\cite{liu_visualizing_2011, Gierz2012,hwang2011}. We thus measure energy- and momentum-resolved photoemission intensities while continuously modulating the XUV polarization state (by rotating a half-wave plate in front of a fixed quarter-wave plate) and extract the CDAD by Fourier-analyzing the polarization-modulated signal~\cite {Fragkos25valley, Fragkos2025} (see Fig.~\ref{FigS1}). The measured dichroism (Fig.~\ref{Fig1}d, right) changes sign between the two Dirac-cone branches and again upon crossing the Dirac energy. These features are consistent with the expected equilibrium pseudospin texture~\cite{Gierz2012}. 

\begin{figure}[H]
\centering
\includegraphics[width=0.9\textwidth]{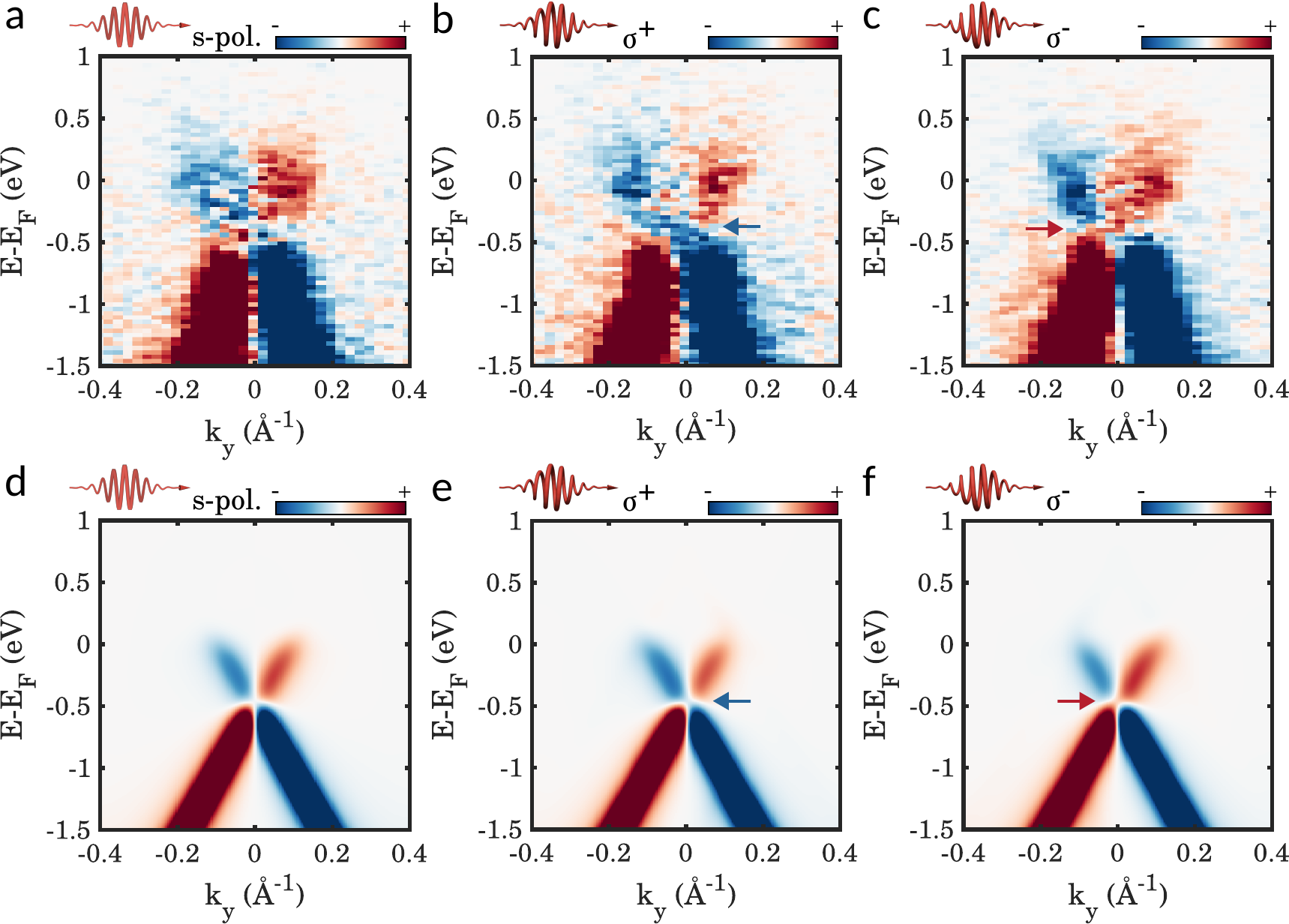}
\caption{\textbf{Pump-polarization-dependent XUV CDAD in light-driven graphene.} \textbf{(a)} Measured CDAD at zero pump-probe delay for a linearly polarized (s-) pump, which retains the sign structure and symmetry of the static CDAD. \textbf{(b)-(c)} Measured CDAD for a right- ($\sigma^+$) and left- ($\sigma^-$) circularly polarized pump, respectively. For both pump helicities, the dichroism acquires a finite contribution close to the Dirac point that is absent for the s-polarized pump. Its sign changes with pump helicity: negative (blue arrow) for $\sigma^+$ and positive (red arrow) for $\sigma^-$. The incident fluence is $\sim$3.5~mJ/cm$^2$, i.e., $\sim$4000~kV/cm. \textbf{(d)-(f)} Corresponding ab initio calculations of the CDAD for graphene driven by \textbf{(d)} s-polarized, \textbf{(e)} $\sigma^+$, and \textbf{(f)} $\sigma^-$ pumps, with a drive amplitude of 4000~kV/cm. In contrast to Fig.~\ref{Fig1}b, these simulations account for substrate effects and associated photoemission matrix elements.}
\label{Fig2}
\end{figure}

We now turn our attention to light-driven graphene. Figure~\ref{Fig2}a–c shows the CDAD at zero pump–probe delay for a linearly polarized (s-polarized) pump and for $\sigma^+$- and $\sigma^-$-circularly polarized pump pulses (incident fluence of $\sim$3.5~mJ/cm$^2$, i.e., $\sim$4000~kV/cm). Upon s-polarized pump, time-reversal symmetry is preserved, leaving the dichroism essentially unchanged from its equilibrium behavior (Fig.~\ref{Fig1}d), with no additional feature emerging around the Dirac point. In contrast, circularly polarized pumping induces an additional dichroic contribution near the Dirac point, whose sign reverses with the pump helicity (see red and blue arrows in Fig.~\ref{Fig2}b-c and e-f). As discussed above, the leading intrinsic contribution to the CDAD is proportional to the out-of-plane pseudospin component $\sigma_z$~\cite{Schueler20}. These new dichroic features therefore reflect the pump-induced modification of the pseudospin texture. Measurements at a lower pump fluence of 1.6~mJ/cm$^2$ (Fig.~\ref{FigS2}) reveal no obvious helicity-dependent dichroic feature near the Dirac point, consistent with the reduced strength of the light-induced mass under weaker driving.

\begin{figure}[H]
\centering
\includegraphics[width=0.9\textwidth]{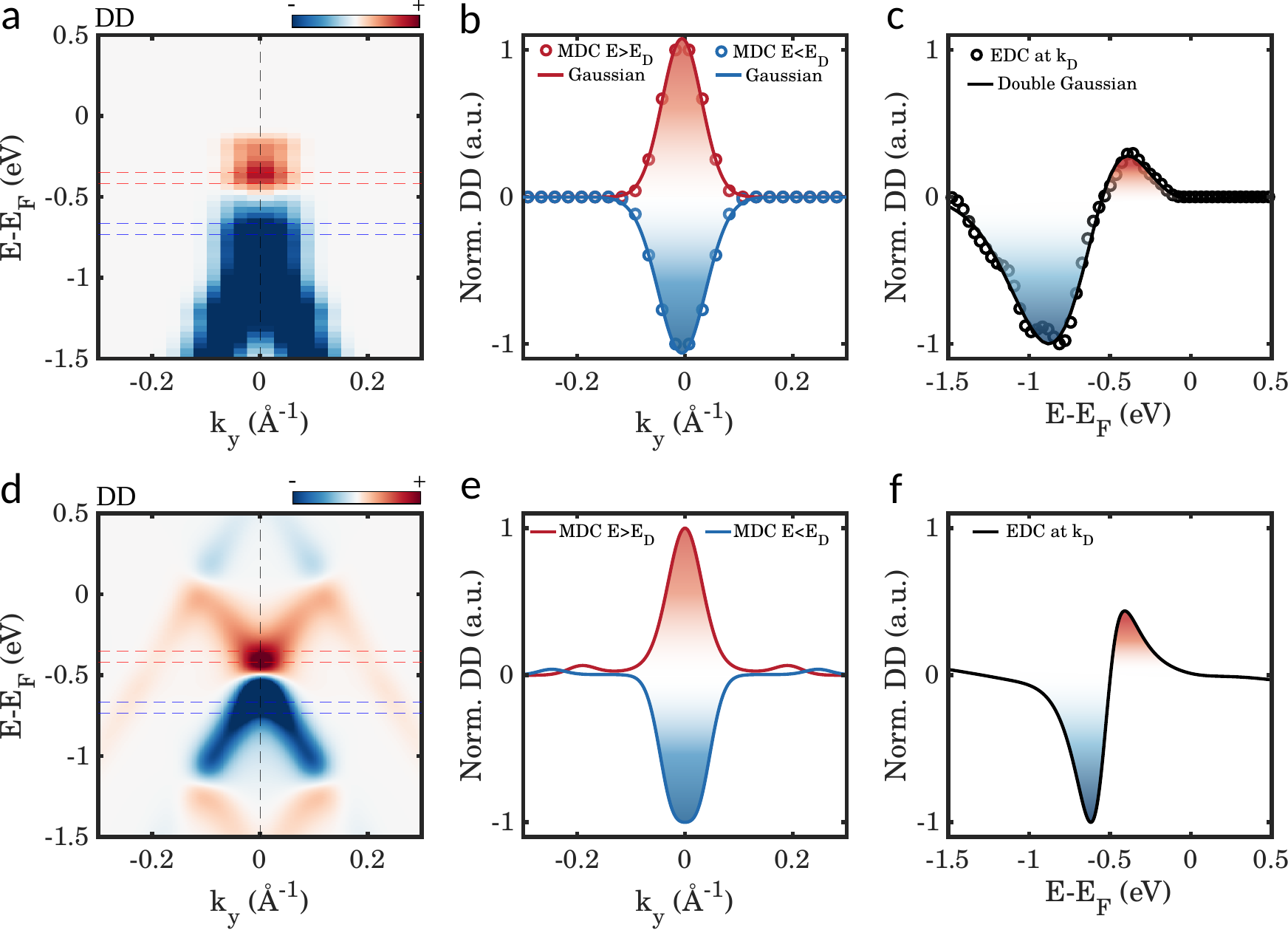}
\caption{\textbf{Spectroscopic signature of light-induced topological state.} \textbf{(a)} Double-helicity dichroism (DD), obtained from the difference between the CDAD measured with $\sigma^+$ and $\sigma^-$ pump (Fig.~\ref{Fig2}b,c), which isolates the pump-helicity-odd component of the dichroism. \textbf{(b)} MDCs of the DD extracted above ($E>E_D$, red) and below ($E<E_D$, blue) the Dirac energy, together with Gaussian fits. \textbf{(c)} EDC of the DD extracted at the Dirac momentum $k_D$, together with a double-Gaussian fit. \textbf{(d-f)} Corresponding ab initio calculations of the DD (d), the MDCs above and below $E_D$ (e), and the EDC at $k_D$ (f), reproducing the localization of the DD around the Dirac point and its sign change across the Dirac energy.}
\label{Fig3}
\end{figure}

The precise energy and momentum dependence of these dichroic features is governed by the photoemission matrix elements. For graphene on SiC at a photon energy of 21.6~eV, matrix-element effects strongly redistribute the spectral weight between the lower and upper halves of the Dirac cone (Fig.~\ref{FigS3}). As a result, in contrast to free-standing graphene (Fig.~\ref{Fig1}b), the light-induced dichroic contributions do not exhibit a direct sign reversal across the Dirac energy. Instead, they appear as a single-sign feature: positive (red) for $\sigma^-$ pumping and negative (blue) for $\sigma^+$ pumping. Our calculations capture this asymmetry (see Methods) and reproduce the measured CDAD for all three pump polarizations (Fig.~\ref{Fig2}d–f). The opposite signs of the dichroic response above and below the Dirac point will, however, become evident in the DD(Fig.~\ref{Fig3}), as discussed below.

To reveal the sign change of the dichroic response across the Dirac energy, we use the doubly differential observable DD (Fig.~\ref{Fig3}), which removes the equilibrium dichroism and other contributions that are independent of the pump helicity, thereby isolating the helicity-odd response of the driven system. As shown in Fig.~\ref{Fig3}a, the measured DD signal is positive above the Dirac energy and negative below it. This behavior is further highlighted by the normalized momentum distribution curves (MDCs) extracted from DD above and below the Dirac energy (Fig.~\ref{Fig3}b), which exhibit a positive peak for $E>E_D$ and a negative dip for $E<E_D$. The corresponding energy distribution curve (EDC), extracted at the Dirac momentum (Fig.~\ref{Fig3}c), further confirms this sign reversal of DD across the Dirac energy.

The sign reversal of DD is consistent with the opposite signs of the light-induced out-of-plane pseudospin in the two bands. Our ab initio calculations (Fig.~\ref{Fig3}d–f), which account for both the Floquet dressing of the initial states and the photoemission matrix elements, reproduce the measured DD. Specifically, they capture the opposite signs of the MDCs above and below $E_D$ (Fig.~\ref{Fig3}e), and the sign reversal of the EDC at the Dirac momentum $k_D$ (Fig.~\ref{Fig3}f). This agreement supports the interpretation of the observed DD as a signature of the light-induced modification of the pseudospin texture of the Floquet topological insulating states. 

\subsection{Link between the light-induced pseudospin texture and Berry curvature}

Here, we go one step further by analyzing the link between the light-induced pseudospin texture, Berry curvature, and CDAD. The different momentum dependencies of the pseudospin and Berry curvature can be understood from the effective Floquet Hamiltonian $H_{\mathrm{eff}}(\mathbf{k})=\mathbf{D}(\mathbf{k})\cdot \hat{\boldsymbol{\sigma}}$, where $\hat{\boldsymbol{\sigma}}$ is the vector of Pauli matrices acting on the A/B sublattice space and $\mathbf{D}(\mathbf{k})$ is the Dirac vector. The eigenenergies are $\pm|\mathbf{D}(\mathbf{k})|$, and the pseudospin expectation values of the two bands are aligned and anti-aligned with $\mathbf{D}(\mathbf{k})$, respectively. Near the Dirac points, the out-of-plane component of $\mathbf{D}$ is given by the light-induced mass $\pm m_F$, with opposite signs at the two valleys. The out-of-plane pseudospin is therefore proportional to $m_F/|\mathbf{D}(\mathbf{k})|$, where $|\mathbf{D}(\mathbf{k})|=\sqrt{(\hbar v_F k)^2+m_F^2}$, $v_F$ is the Fermi velocity, and $\mathbf{k}$ is measured relative to the respective $K$ or $K^\prime$ point. The Berry curvature of the two bands is given by $\Omega_\pm(\mathbf{k}) = \mp \frac{1}{2}\hat{\mathbf{D}}\cdot (\partial_{k_x}\hat{\mathbf{D}}\times \partial_{k_y}\hat{\mathbf{D}})$, where $\hat{\mathbf{D}}=\mathbf{D}/|\mathbf{D}|$.

\begin{figure}[H]
\centering
\includegraphics[width=0.9\textwidth]{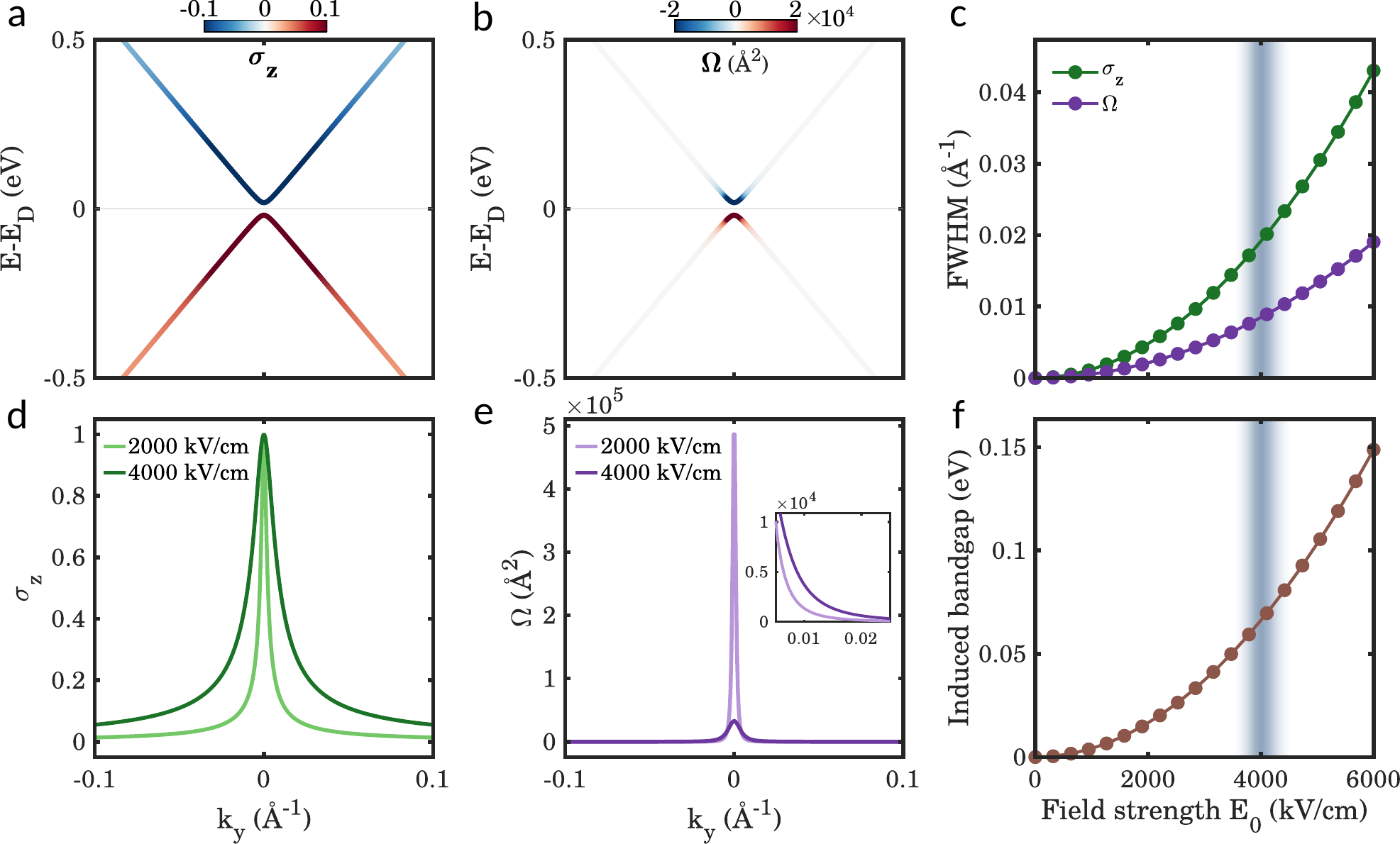}
\caption{\textbf{Link between the light-induced pseudospin and the Berry curvature.} \textbf{(a,b)} Band-resolved out-of-plane pseudospin expectation value $\sigma_z$ (a) and Berry curvature $\Omega$ (b), overlaid on the Floquet band structure along $k_y$ for a drive amplitude of 3000~kV/cm. \textbf{(c)} Full width at half maximum (FWHM) of $\sigma_z$ and $\Omega$ as a function of the field strength. \textbf{(d,e)} Momentum profiles of $\sigma_z$ (d) and $\Omega$ (e) for field strengths of 2000 and 4000~kV/cm. The inset in (e) shows the tails of $\Omega$. \textbf{(f)} Induced band gap as a function of the field strength. Calculations use the Floquet Hamiltonian of a tight-binding model of graphene (see Methods). Gray areas in (c) and (d) highlight the experimental field strength.}
\label{Fig4}
\end{figure}

Figure~\ref{Fig4}a,b compares the band-resolved out-of-plane pseudospin $\sigma_z(\mathbf{k})$ and Berry curvature $\Omega(\mathbf{k})$, calculated from the Floquet Hamiltonian of the tight-binding model. Within the massive-Dirac description, their magnitudes vary as $|\sigma_z(\mathbf{k})| \propto |m_F|[(\hbar v_F k)^2+m_F^2]^{-1/2}$ and $|\Omega(\mathbf{k})| \propto |m_F|(\hbar v_F)^2[(\hbar v_F k)^2+m_F^2]^{-3/2}$. Away from the Dirac point, the Berry curvature thus decays as $1/k^3$, whereas the pseudospin decays only as $1/k$. This difference is illustrated in Fig.~\ref{Fig4}c-e. While the Berry curvature is strongly localized near the Dirac point, the pseudospin extends further along the bands. This is particularly useful for small light-induced gaps (Fig.~\ref{Fig4}f), where scattering can obscure the gap without completely suppressing the dichroic signal.

\section*{Conclusions}

Our results provide direct spectroscopic evidence for a Floquet topological insulating state in graphene, finally validating a landmark theoretical prediction that has long driven efforts to realize light-induced topological phases of matter. Rather than inferring topology indirectly from spectral signatures, our helicity-resolved photomission circular dichroic measurements directly access the quantum geometry of light-dressed wavefunctions. We resolve the pseudospin texture generated by the light-induced Haldane mass in excellent agreement with ab initio theory and simulations. These findings establish momentum-resolved wavefunction geometry as a direct probe of Floquet topology, robust even in the presence of scattering and heating. Furthermore, pump helicity controls the signs of the induced mass, Berry curvature, and orbital magnetization, enabling reversal of these properties on the optical-pulse timescale. This capability opens a route to writing, switching, and erasing topological states with femtosecond temporal resolution. Our results thus establish topology not merely as a static property of a quantum ground state, but as a dynamically controllable degree of freedom, opening new opportunities for ultrafast topological control in quantum materials.

\begin{methods}
\subsection{Time- and polarization-resolved momentum microscopy} 

The experiments were carried out using a polarization-controllable ultrafast extreme ultraviolet (XUV) beamline~\cite{Comby22} connected to a momentum microscope end-station~\cite{tkach24, tkach24-2, Fragkos2025} at the Centre Lasers Intenses et Applications (CELIA). The laser source was a high-repetition-rate Yb fiber laser (166~kHz, 1030~nm, 135~fs, 50~W; Amplitude Laser Group). In the high-harmonic generation (HHG) arm, a substantial fraction of the fundamental laser output was frequency-doubled in a BBO crystal, yielding several watts of 515~nm radiation. The resulting beam was resized and spatially shaped into an annular profile before being focused into an argon gas jet, to generate XUV radiation through HHG. After the gas jet, residual 515~nm radiation was suppressed through successive spatial filtering stages employing pinholes. The ninth harmonic of the 515~nm driving field, corresponding to a photon energy of 21.6~eV, was selected using reflections from Sc/SiC multilayer XUV mirrors (NTTAT) in combination with a 200~nm-thick Sn filter (Luxel).

The infrared (IR) pump pulses were obtained by extracting a small fraction of the fundamental laser output (1030~nm, 135~fs). The polarization was adjusted using a motorized quarter-wave plate positioned immediately upstream of the recombination chamber. The pump and probe beams were combined collinearly using a drilled mirror and focused onto the sample. The overall temporal resolution was approximately 150~fs. The IR pump fluence was determined in situ using the photoemission electron microscopy (PEEM) mode of our time-of-flight momentum microscope, which enables real-space imaging of photoelectrons emitted from the sample surface. Because photoemission induced by the 1.2~eV IR pulses is a multiphoton process, the measured PEEM image does not directly reproduce the spatial profile of the IR beam. Instead, it reflects the beam profile modified by the effective nonlinearity of the photoemission process. To account for this effect, we determined the effective nonlinearity by measuring the total photoemission intensity as a function of the IR pump power. Using the extracted nonlinearity, we reconstructed the actual IR beam profile from the PEEM images, enabling an accurate in situ determination of the pump fluence.

The graphene sample (Graphene-SiC) was mounted on a motorized six-axis hexapod, which enabled precise sample positioning and alignment in the main chamber. Measurements were conducted at room temperature under a base pressure of 2$\times10^{-10}$~mbar. Time-resolved photoemission data were acquired with a custom-built time-of-flight momentum microscope featuring an advanced front lens with multiple operating modes (GST mbH)~\cite{tkach24, tkach24-2}. For XUV circular dichroism in the angular distribution (CDAD) measurements, an all-reflective XUV quarter-wave plate was placed in the probe beam path. During acquisition, a visible-light half-wave plate (515~nm), positioned upstream of the HHG chamber (before the XUV quarter-wave plate), was continuously rotated while the photoemission signal was recorded. Each detected electron event was correlated with the instantaneous angular position of the half-wave plate. Owing to the reflection of the XUV beam from the quarter-wave plate, one complete 360$^{\circ}$ rotation of the driving-beam half-wave plate generated four successive cycles of p-, left-circularly polarized (LCP), s-, right-circularly polarized (RCP), and p-polarized XUV radiation. To determine the dichroic response, we performed a Fourier analysis independently for each energy-momentum voxel, resolving the photoemission intensity as a function of the XUV polarization angle. The real and imaginary components of the resulting oscillatory signal were used to quantify the linear and circular dichroism in XUV photoemission, respectively. Raw single-event data were processed using an open-source data-analysis workflow~\cite{Xian20, Xian19_2}, which converts the event-level measurements into calibrated, binned hypervolumes with the required dimensions. Further details of the experimental apparatus and the CDAD extraction using Fourier analysis are provided in Ref.~\cite{Fragkos2025}.

\subsection{Graphene sample} 

The \(7~\mathrm{mm}\times7~\mathrm{mm}\) monolayer epitaxial graphene sample on SiC was purchased from Graphene-SiC. The graphene was synthesized by graphitization of the Si-face of 4H–SiC~\cite{Emtsev2009}. The sample is predominantly covered by monolayer graphene, with additional bilayer graphene present in the form of stripe-like domains. According to the manufacturer, optical inspection confirmed a monolayer coverage exceeding \(95\%\), with bilayer graphene inclusions at \(5\%\). The sample was annealed at 400$^\circ$C for 2~hours to prepare a clean surface before photoemission measurements.  

\subsection{Theoretical calculations}

The simulations of the trARPES signal at pump-probe overlap were carried out using the non-equilibrium Green's function formalism, specifically the Floquet steady-state approach described in ref.~\cite{Schuler_WSe2}. In brief, the photoemission intensity is computed from
\begin{align}
    I(\mathbf{k},E) = \mathrm{Im}\sum_{nn'} M_n(\mathbf{k},E) G^<_{nn'}(\mathbf{k}, \omega_\mathrm{probe} - E) M^*_{n'}(\mathbf{k},E) \ ,
\end{align}
where $M_n(\mathbf{k},E)$ are the photoemission matrix elements, and where $G^<_{nn'}(\mathbf{k},\omega)$ is the period-averaged lesser Green's function. The kinetic energy of the photoelectrons is denoted by $E$, while $\omega_\mathrm{probe}$ is the probe photon energy. To calculate the Green's function, two ingredients are required: (i) the light-matter coupling of the pump pulse, and (ii) a dissipative bath defining the steady state and the effective occupations. The latter is described by a wide-band fermionic bath with a coupling strength of $\gamma = 0.25$~eV for the conduction band and $\gamma = 0.125$~eV for the valence band. A chemical potential of $\mu = 0.5$~eV s used to emulate the doping of the sample. For (i), we computed the velocity matrix elements $\mathbf{v}_{nn'}(\mathbf{k}) = i [\hat{H}_\mathrm{KS}(\mathbf{k}), \hat{\mathbf{r}}]_{nn'}$ directly from the Kohn-Sham (KS) Hamiltonian after performing density-functional theory (DFT) calculations of the band structure. In the velocity gauge, this procedure yields the time-dependent Hamiltonian
\begin{align}
    \label{eq:td_ham}
    H_{nn'}(\mathbf{k}, t) = \left( \varepsilon_n(\mathbf{k}) + \frac{q^2 \mathbf{A}(t)^2}{2}\right)\delta_{nn'} - q \mathbf{A}(t)\cdot \mathbf{v}_{nn'}(\mathbf{k})  \ ,
\end{align}
where $\mathbf{A}(t) = (E_0 / \omega_\mathrm{pump}) \mathrm{Re}[i \hat{u} e^{-i\omega_\mathrm{pump} t}]$ is the pump vector potential. The effective polarization vector $\hat{u}$ was computed as the superposition of the incoming and reflected fields at the interface of the SiC substrate under the experimental angle of incidence, using a dielectric constant of $\epsilon_\mathrm{SiC}(\omega_\mathrm{pump}) = 6.7$. The time-dependent Hamiltonian~\eqref{eq:td_ham} is then used to construct the Floquet Hamiltonian and $G^<_{nn'}(\mathbf{k}, \omega)$ (for details see ref.~\cite{Schuler_WSe2}). We also include laser-assisted photoemission (LAPE) in the Green's function.

The photoemission matrix elements $M_n(\mathbf{k},E)$ for free-standing graphene are computed from first principles using the approach described in ref.~\cite{parusa_accurate_2026}. Compared to the experiments, we notice some deviations in the equilibrium ARPES signal, which are induced by the SiC substrate. In particular, the photoemission intensity is strongly reduced for p-polarization; in the experimental geometry, this is only possible for an almost vanishing dielectric constant $\epsilon_\mathrm{SiC}(\omega_\mathrm{probe})$. The most likely explanation is a bulk plasmon resonance, which has been observed in electron energy-loss spectroscopy experiments~\cite{yang_electron_2020} at $\sim 22$~eV. A simple density counting argument also yields $\omega_p = \sqrt{4\pi n} \simeq 22$~eV. We fix $\epsilon_\mathrm{SiC}(\omega_\mathrm{probe}) = 0.02 + 0.1i$ and compute the probe polarization from the Fresnel equations (superposition of incoming and reflected fields). Additionally, we scale the photoemission matrix element for p-polarization, $M^p_c(\mathbf{k}, E) \rightarrow r M^p_c(\mathbf{k}, E)$ $r = \frac{1}{\sqrt{2}}$. To account for the reduced intensity of circular dichroism results above the Dirac point, we further scale all intensities with a smooth step function going from 1 below $E_D$ to 1/2 above $E_D$.This model for matrix elements yields polarization-resolved equilibrium spectra in good agreement with the experiments (see Fig.~\ref{FigS3}), including the circular dichroism.

For the results presented in Fig.~\ref{Fig4}, we used the two-band Dirac model $H_D(\mathbf{k}) = \mathbf{D}(\mathbf{k})\cdot \hat{\boldsymbol{\sigma}}$, where $\mathbf{D}(\mathbf{k}) = (-\tau\hbar v_F k_x, \hbar v_F k_y, 0 )$. Here, $\tau$ is the valley index, while $v_F$ denotes the Fermi velocity.
We then constructed the time-dependent Hamiltonian $H_D(\mathbf{k},t) = H_D(\mathbf{k} - q\mathbf{A}(t))$ via the Peierls substitution and constructed the corresponding Floquet Hamiltonian $\mathcal{H}_{am,a'm'}(\mathbf{k})$, where $a,a'$ ($m,m'$) denote the sublattice sites (Floquet indices). Diagonalizing the Floquet Hamiltonian yields the quasienergies $\varepsilon_\lambda(\mathbf{k})$ with corresponding eigenvectors $W_{am,\lambda}(\mathbf{k})$, from which we compute the pseudospin $\sigma_z(\mathbf{k}) = |W_{Am,\lambda}(\mathbf{k})|^2 - |W_{Bm,\lambda}(\mathbf{k})|^2$ for band and Floquet indices $\lambda, m$ corresponding to the effective valence band. Similarly, the Berry curvature is computed via the sum-over-bands definition Curvature 
\begin{align}
    \label{eq:berryc}
    \Omega_\lambda(\mathbf{k}) = 2\text{Im}\left(\sum_{\lambda'\ne \lambda} \frac{v_{\lambda \lambda', x}(\mathbf{k}) v_{\lambda'\lambda, y}(\mathbf{k}) - v_{\lambda \lambda', y}(\mathbf{k}) v_{\lambda'\lambda, x}(\mathbf{k})}{(\varepsilon_{\lambda'}(\mathbf{k}) - \varepsilon_{\lambda}(\mathbf{k}))^2}\right) \ ,
\end{align}
where $v_{\lambda \lambda',\mu}(\mathbf{k}) = \sum_{m,aa'}W^*_{am,\lambda}(\mathbf{k})v_{aa',\mu}(\mathbf{k})W_{a'm,\lambda}(\mathbf{k})$. The velocity matrix elements $v_{aa',\mu}(\mathbf{k})$ are calculated from the Dirac model here.

\end{methods}

\section*{Data Availability}
The data that support the findings of this article will be openly available on Zenodo before publication \url{}. 

\section*{Acknowledgement} 
We thank Charlotte Sanders for insightful discussions at the project's initial stage, and Michael A. Sentef for fruitful discussions during the final phase. We thank Nikita Fedorov, Romain Delos, Pierre Hericourt, Rodrigue Bouillaud, Laurent Merzeau, and Frank Blais for technical assistance.  We thank Baptiste Fabre for implementing and maintaining the data binning code. We acknowledge the financial support of the IdEx University of Bordeaux/Grand Research Program ``GPR LIGHT". This work is part of the ULTRAFAST and TORNADO projects of PEPR LUMA and was supported by the French National Research Agency, as a part of the France 2030 program, under grants ANR-23-EXLU-0002 and ANR-23-EXLU-0004. We acknowledge support from ERC Starting Grant ERC-2022-STG No.101076639, Quantum Matter Bordeaux, AAP CNRS Tremplin, and AAP SMR from Université de Bordeaux. S.M. and W.B. acknowledge support from the Deutsche Forschungsgemeinschaft (DFG, German Research Foundation), project number 567214325, and by the Lower Saxony Ministry of Science and Culture (MWK) and the VolkswagenStiftung through the zukunft.niedersachsen program "Wissenschaftsräume" (Wissenschaftsraum ElLiKo, "Elektronen–Licht–Kontrolle"). W.B. acknowledges support from the Dorothea Schlözer Postdoctoral Programme for Women. S.F. acknowledges funding from the European Union’s Horizon Europe research and innovation programme under the Marie Skłodowska-Curie 2024 Postdoctoral Fellowship No 101198277 (TopQMat). U.D. acknowledges funding from the European Union’s Horizon Europe research and innovation programme under the Marie Skłodowska-Curie grant No HORIZON-MSCA-2023-DN-01 101169225 - SPARKLE. A.C.G. and M.S. acknowledge funding from the Swiss National Science Foundation (SNSF), grant no. 10003275. P.H. acknowledges support by Novo Nordisk Foundation (Grant number NNF23OC0085585). Funded by the European Union. Views and opinions expressed are, however, those of the author(s) only and do not necessarily reflect those of the European Union. Neither the European Union nor the granting authority can be held responsible for them.

\section*{Author contributions}
S.B. and M.S. conceived the idea. S.F. and S.B. planned and performed the experiments. U.D., W.B. and Y.M. assisted with the experiments. S.F., Y.M. and S.B. analyzed the experimental data. D.D. and S.P. helped maintain the laser system. M.S. and A.C.G. developed the theory and associated interpretations. S.F., S.B., A.C.G. and M.S. provided the initial interpretation of the results. S.F., A.C.G., W.B., P.H., M.R., S.M. and S.B. discussed the results, analysis, and interpretation. S.F., S.B., A.C.G. and M.S. wrote the first draft of the manuscript with input from U.D., D.D., W.B., P.H., M.R., S.M., and Y.M.

\section*{Competing Interests}
The authors declare that they have no competing financial interests.

\clearpage
\bibliographystyle{naturemag}
\bibliography{main}

\clearpage
\begin{center}
{\Large\bfseries\sffamily Supplementary Information:\\ Observation of light-induced topological state in graphene}
\end{center}

\setcounter{figure}{0}
\renewcommand{\thefigure}{S\arabic{figure}}
\setcounter{equation}{0}
\renewcommand{\theequation}{S\arabic{equation}}

\subsection{Simultaneous extraction of CDAD and LDAD.} 
\begin{figure}
\centering
\includegraphics[width=0.95\textwidth]{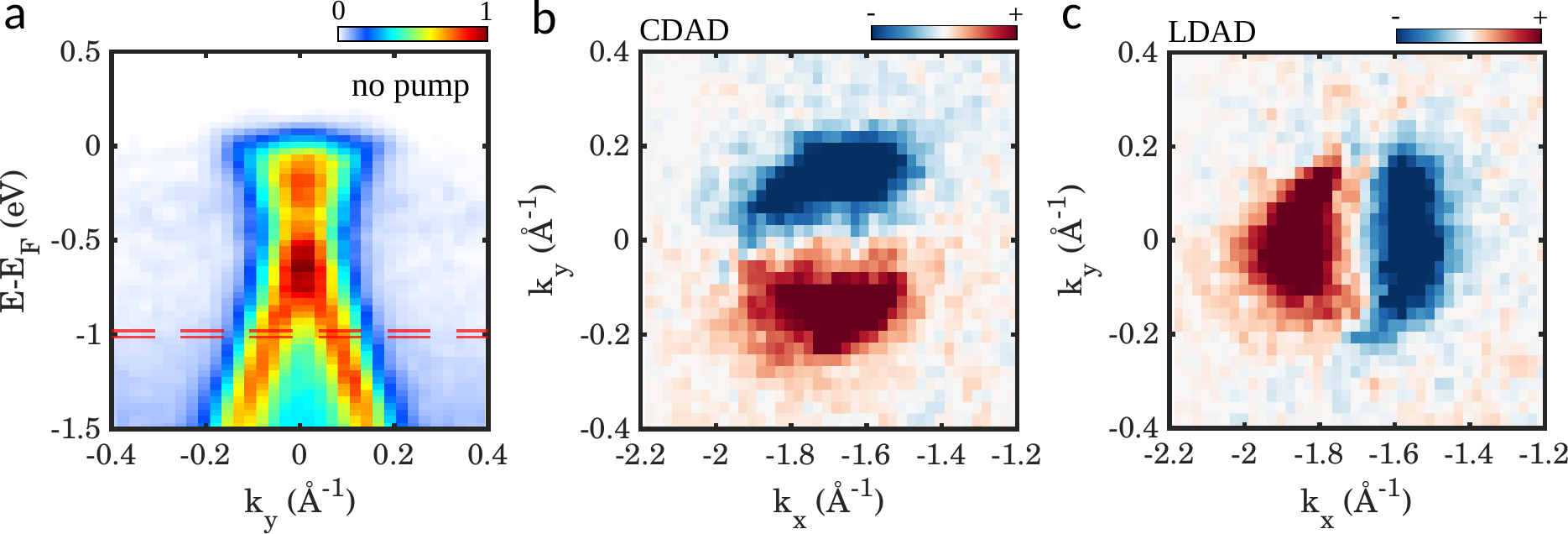}
\caption{\textbf{Simultaneous extraction of CDAD and LDAD.} \textbf{(a)} Photoemission intensity from graphene on SiC as a function of energy and momentum $k_y$ along a cut through the Dirac point, measured without pump excitation. The dashed lines indicate the energy window used to extract the constant-energy contours (CECs) shown in (b,c). \textbf{(b)} CEC of the circular dichroism in the angular distribution (CDAD) around the K point, obtained from the imaginary parts of the Fourier coefficients of the periodically modulated photoemission signal recorded during continuous modulation of the XUV polarization state (see Methods). The CDAD exhibits antisymmetry with respect to $k_y=0$. \textbf{(c)} CEC of the linear dichroism in the angular distribution (LDAD), obtained from the real parts of the Fourier coefficients of the same periodically modulated signal (see Methods). The LDAD exhibits antisymmetry about the K point along $k_x$. Both CECs display the symmetry patterns expected from the pseudospin texture of the Dirac cone, demonstrating that CDAD and LDAD can be reliably extracted simultaneously from a single measurement.}
\label{FigS1}
\end{figure}

The static CDAD and LDAD of graphene shown in Fig.~\ref{FigS1} are extracted from a single measurement in which the XUV polarization state is continuously modulated (see Methods). This single-scan approach avoids systematic errors that can arise when CDAD and LDAD are measured sequentially under nominally identical conditions using separately acquired datasets for different polarization states. It underpins all static and time-resolved dichroism measurements presented in the main text. The corresponding constant-energy contours (Fig.~\ref{FigS1}b,c) exhibit the antisymmetry expected from the pseudospin texture of the Dirac cone~\cite{Schuler20}, confirming the reliable and simultaneous extraction of both observables from the same dataset.

\newpage

\subsection{Fluence dependence of the pump-helicity-dependent dichroism.} 

\begin{figure}[H]
\centering
\includegraphics[width=0.75\textwidth]{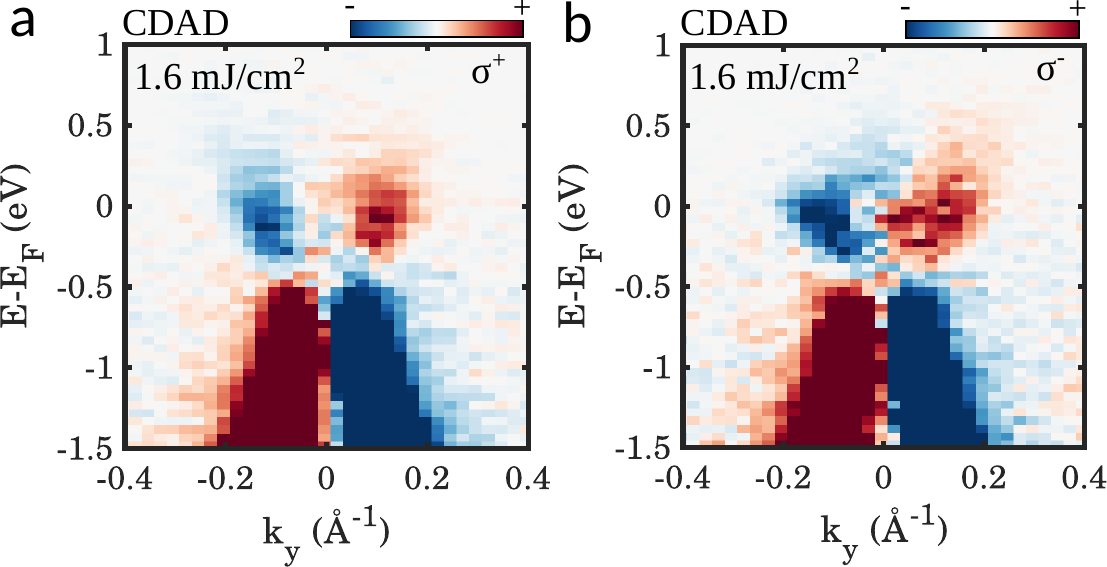}
\caption{\textbf{Circular dichroism at low pump fluence.} Circular dichroism in the angular distribution (CDAD) of graphene as a function of energy and momentum $k_y$ along a cut through the Dirac point, measured at temporal overlap of the pump and probe pulses for a \textbf{(a)} right- ($\sigma^+$) and a \textbf{(b)} left-circularly polarized ($\sigma^-$) pump, at a reduced pump fluence of 1.6~mJ/cm$^2$. No discernible pump-helicity dependence of the CDAD is observed near the Dirac point.}
\label{FigS2}
\end{figure}

We investigate the CDAD at a reduced pump fluence of 1.6~mJ/cm$^2$, compared with the 3.5~mJ/cm$^2$ used in the main text (Fig.~\ref{FigS2}). At this lower fluence, no discernible difference is observed between the CDAD measured using $\sigma^+$ and $\sigma^-$ pump pulses near the Dirac point, in contrast to the pronounced helicity dependence observed in Fig.~2 of the main text. This behavior is consistent with the reduced magnitude of the light-induced mass term at weaker driving strengths. Since this term scales quadratically with the pump field amplitude, its effect is expected to fall below our detection sensitivity at sufficiently low pump fluences.

\newpage

\subsection{XUV polarization dependence of the photoemission intensities}

\begin{figure}[H]
\centering
\includegraphics[width=0.95\textwidth]{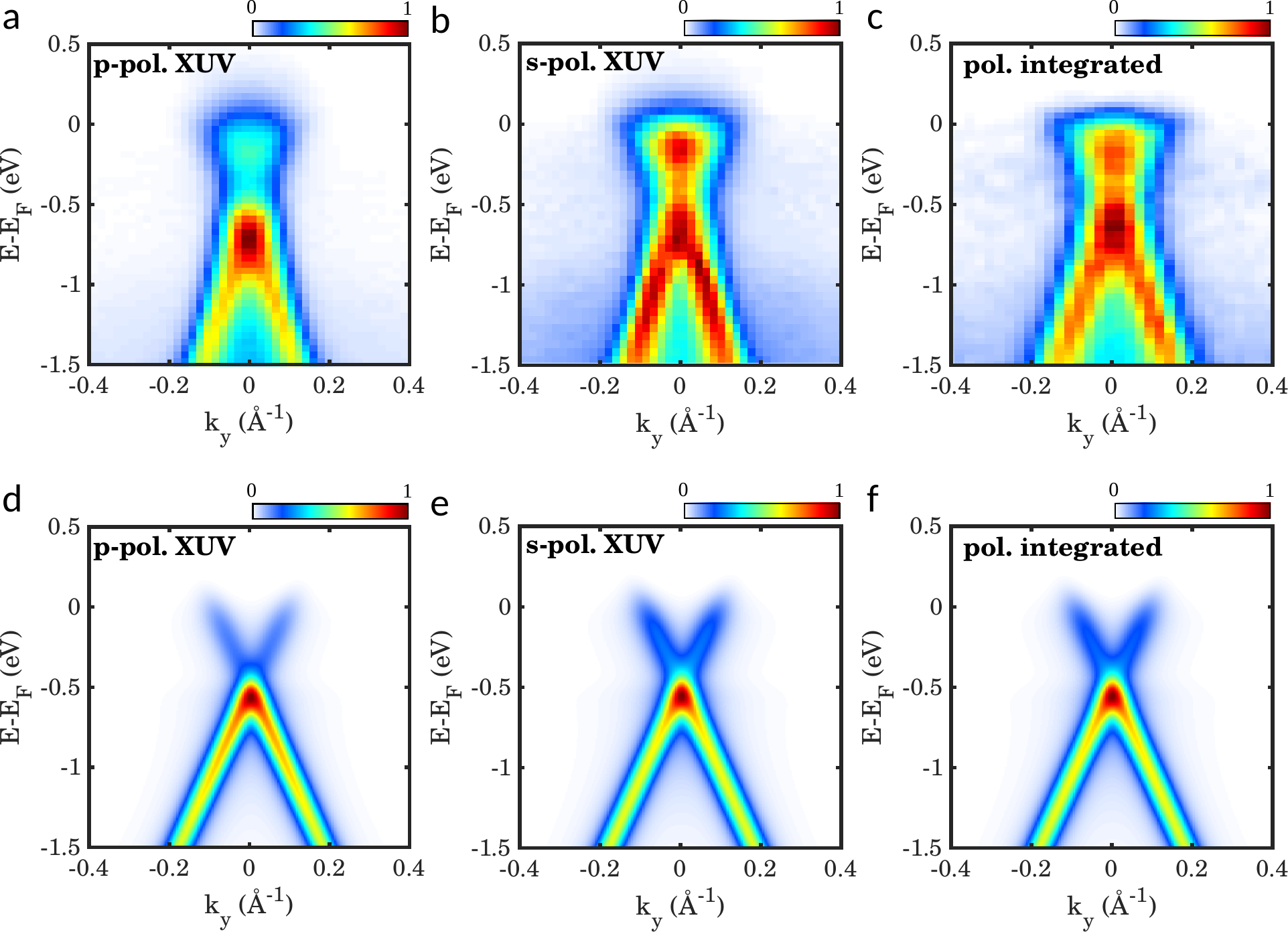}
\caption{\textbf{Polarization dependence of the photoemission intensity.}
\textbf{(a-c)} Measured photoemission intensity of graphene on SiC along $k_y$ through the Dirac point at a photon energy of 21.6~eV, with \textbf{(a)} p-polarized XUV, \textbf{(b)} s-polarized XUV, and \textbf{(c)} integrated over all polarization states during continuous waveplate rotation (see Methods). In (a) and (b), the intensity is normalized to its maximum. In (c), the intensity at each QWP angle is normalized to its maximum in every energy row before integration, which reduces the matrix-element contribution. \textbf{(d-f)} Corresponding calculated photoemission intensity for p-polarized XUV (d), s-polarized XUV (e), and XUV integrated over all polarization states, including linear and circular polarizations (f). Each calculated map is normalized to its own maximum.}
\label{FigS3}
\end{figure}

Fig.~\ref{FigS3} shows the photoemission intensity of graphene on SiC along $k_y$ through the Dirac point, recorded at a photon energy of 21.6~eV for different XUV probe polarizations. The distribution of spectral weight between the upper and lower Dirac cones depends strongly on the probe polarization. For p-polarized XUV (Fig.~\ref{FigS3}a), the intensity above the Dirac point is strongly suppressed, with most of the spectral weight concentrated in the lower cone. For s-polarized XUV (Fig.~\ref{FigS3}b), the intensity is distributed more evenly between the two cones, and both branches remain visible across the Dirac point. Since circularly polarized XUV contains a p-polarized component, a similar imbalance is expected in the CDAD measurements. This is consistent with the predominantly single-sign CDAD observed near the Dirac point (Fig.~2), in contrast to the calculations for free-standing graphene (Fig.~1b). The observed behavior is reproduced by the calculations when the modified photoemission matrix elements are taken into account.

The suppression of the upper cone also leads to an intensity depletion near the Dirac point that resembles a gap. This feature is substantially more pronounced for p-polarized than for s-polarized XUV, indicating a strong influence of photoemission matrix elements. To mitigate these effects, we integrate the photoemission intensity over the range of polarization states sampled during continuous rotation of the half-wave plate placed before the XUV quarter-wave plate (see Methods). Before integration, the intensity recorded at each QWP angle is normalized to its maximum within each energy row. This procedure reduces the polarization- and energy-dependent intensity modulation introduced by the photoemission matrix elements while preserving the band dispersion. In the resulting spectrum (Fig.~\ref{FigS3}c), the intensity distribution becomes more continuous across the Dirac point, and the gap-like depletion is further suppressed. We therefore attribute this feature primarily to photoemission matrix-element effects, although a small gap below our experimental resolution cannot be ruled out.

Our simulations of the equilibrium ARPES spectrum, using the photoemission matrix elements $M_n(\mathbf{k}, E)$ for free-standing graphene and applying the scaling explained in the Methods section, are in good qualitative agreement with the experimental spectra (Fig.~\ref{FigS3}d--f). In particular, the suppressing of the photoemission intensity above $E_D$ for p polarization is reproduced well. We use the thus calibrated photoemission matrix elements consistently for all calculated results in the main text.

\subsection{Effect of a trivial gap on pseudospin, Berry Curvature, and dichroisms.}

In the main text, the pump-helicity-dependent CDAD near the Dirac point is interpreted as a signature of the light-induced pseudospin texture. Since graphene on SiC can also exhibit a trivial gap, we examine whether our main interpretation, namely, that the DD probes the pseudospin texture associated with a Floquet topological insulator, remains robust in the presence of such a trivial gap already present at equilibrium.

For the \emph{ab initio} Hamiltonian, a gap can be introduced by modifying the Hamiltonian  
\begin{align}
\label{eq:eff_ham}
\tilde{H}_{nn'}(\mathbf{k}) = \varepsilon_n(\mathbf{k})\delta_{nn'} + \frac{\Delta}{2i |C(\mathbf{k})|} [v_x(\mathbf{k}), v_y(\mathbf{k})]_{nn'} \ ,
\end{align}
where $|C(\mathbf{k})|$ is the largest eigenvalue of the commutator $[v_x(\mathbf{k}), v_y(\mathbf{k})]$. For the TB model, the gap is introduced by replacing the diagonal elements by $\Delta$ ($-\Delta$) for the A (B) sublattice site. In both models, the system exhibits a gap $E_\mathrm{gap}=|\Delta|$ and a trivial mass of $|\Delta|/2$. The sign of $\Delta$ is determined by the spatial arrangement of the graphene sheet on the SiC substrate. Fixing a value for $\Delta$, a net CDAD is observed even in equilibrium, which is not observed in the experiments. Thus, we can assume that domains with different orientation are present, with roughly equal contributions to the total photoemission signal. Mathematically, this scenario is modeled by performing calculations for $\pm \Delta$ and averaging the results. Eq.~\eqref{eq:eff_ham} also allows us to study pure Floquet effects within the high-frequency perturbation theory by identifying $\Delta$ with the Floquet mass $m_F$.

\begin{figure}[H]
\centering
\includegraphics[width=0.7\textwidth]{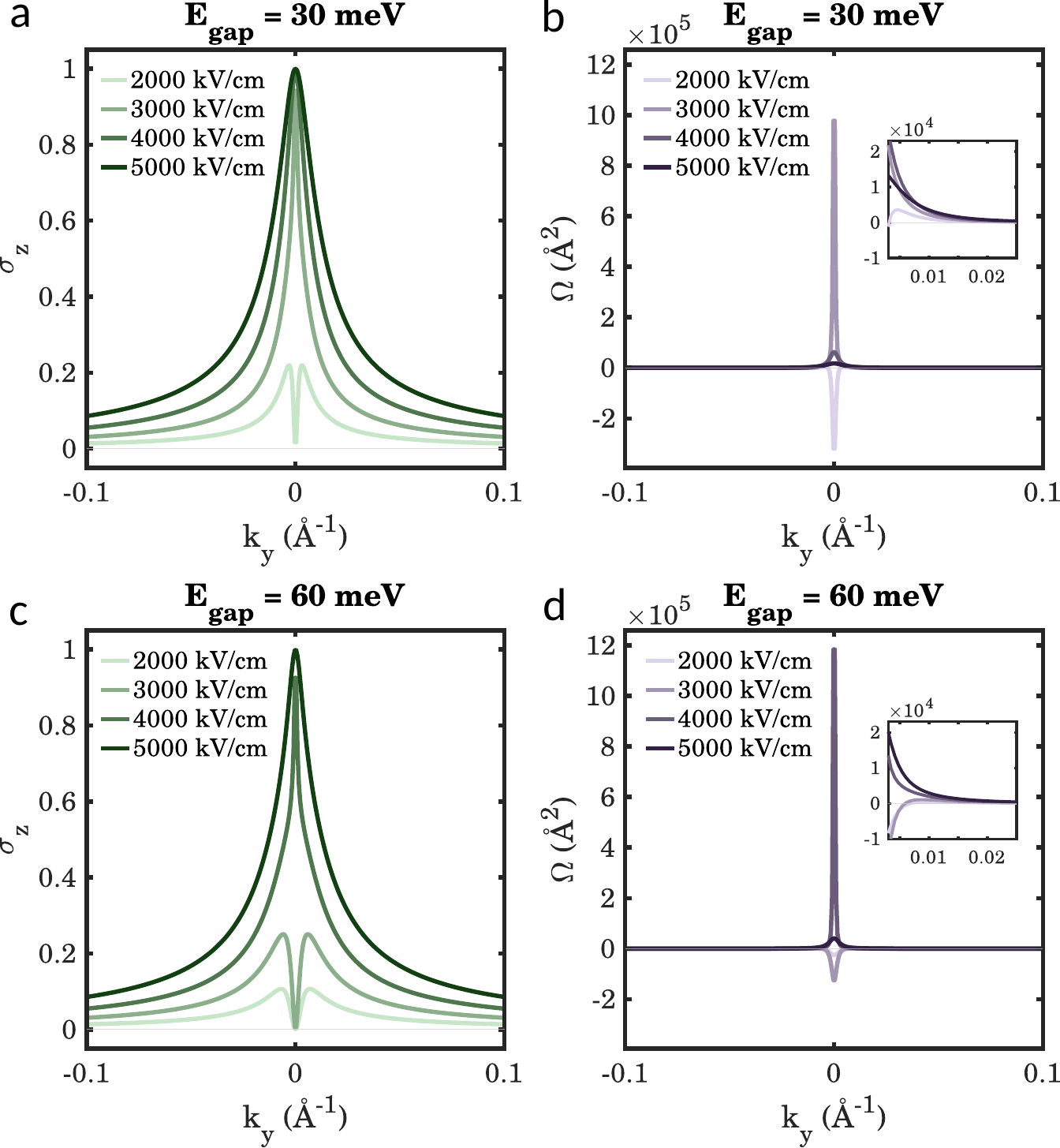}
\caption{\textbf{Impact of trivial substrate-induced gap.} Light-induced pseudospin $\sigma_z(\mathbf{k})$ and Berry curvature for the gapped model, for $E_\mathrm{gap}=30$~meV  (\textbf{a,b}) and $E_\mathrm{gap}=60$~meV  (\textbf{c,d}). The result were obtained from the high-frequency expansion of the TB model. \label{fig:pseudo_bc_gap}}
\end{figure}

Fig.~\ref{fig:pseudo_bc_gap} shows the light-induced pseudospin and Berry curvature in the presence of the trivial gap. For small field strength, the pseudospin is pinned to zero at $k_y = 0$ relative to the Dirac point; away for $k_y=0$, the behavior closely resembles the case of $E_\mathrm{gap} = 0$. As soon as the light-induced gap exceeds the trivial one, pseudospin approaches 1, and the profile of $\sigma_z(\mathbf{k})$ becomes very similar to the pristine graphene case (Fig.~\ref{fig:pseudo_bc_gap}a,c). The Berry curvature exhibits similar behavior (Fig.~\ref{fig:pseudo_bc_gap}b,d).

\begin{figure}[H]
\centering
\includegraphics[width=0.95\textwidth]{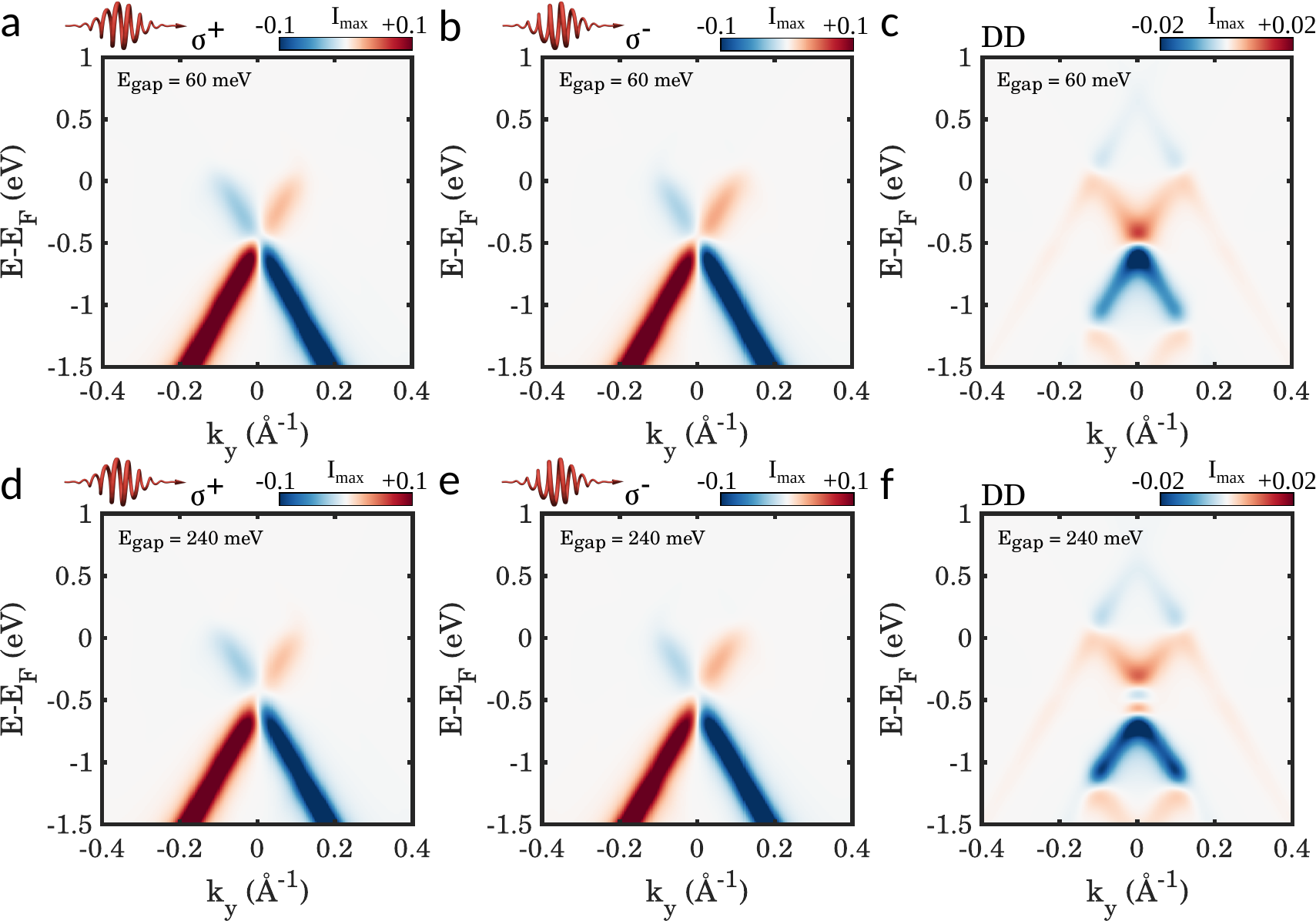}
\caption{\textbf{Effect of trivial band gap on CDAD and DD.}
Calculated XUV CDAD along $k_y$ through the Dirac point for pump helicity $\sigma^{+}$ (\textbf{a},\textbf{d}) and $\sigma^{-}$ (\textbf{b},\textbf{e}), and the corresponding double difference (DD) between the two pump helicities (\textbf{c},\textbf{f}). Panels \textbf{a--c} correspond to a trivial gap $E_\mathrm{gap}=60$~meV and panels \textbf{d--f} to $E_\mathrm{gap}=240$~meV, for a drive field of 4000~kV/cm. The dichroic features near the Dirac point, associated with the pseudospin and Berry-curvature texture, are present for both gap values, but are fading and accompanied by a large region of vanishing signal in the DD for the larger gap.}
\label{FigS4}
\end{figure}

Next, we investigated the impact of the substrate-induced gap onto the photoemission signal and the circular dichroism.
Figure~\ref{FigS4} shows the calculated CDAD maps and corresponding DD for $E_\mathrm{gap}=60$~meV (a--c) and $240$~meV (d--f), both for the peak field strength of $E_0 = 4000$~kV/cm used for all simulations in the main text. For the the smaller  gap of $E_\mathrm{gap} = 60$~meV, the CDAD and the DD are very similar to the pristine case, consistent with behavior the light-induced pseudospin and Berry curvature (Fig.~\ref{fig:pseudo_bc_gap}a,c). These results show that the dichroic features associated with the pseudospin and Berry curvature do not require the trivial equilibrium gap to vanish.
Increasing the gap beyond the Floquet gap, 
the dichroic signal near the Dirac point is visibly suppressed in the CDAD maps (Fig.~\ref{FigS4}d,e), while the DD (Fig.~\ref{FigS4}f) develops an extended region with a vanishing signal. Neither behavior is observed in our experimental data, suggesting that our samples do not lie in the regime of such a large trivial gap.

\subsection{Topological origin of the dichroism}

\begin{figure}[H]
\centering
\includegraphics[width=0.75\textwidth]{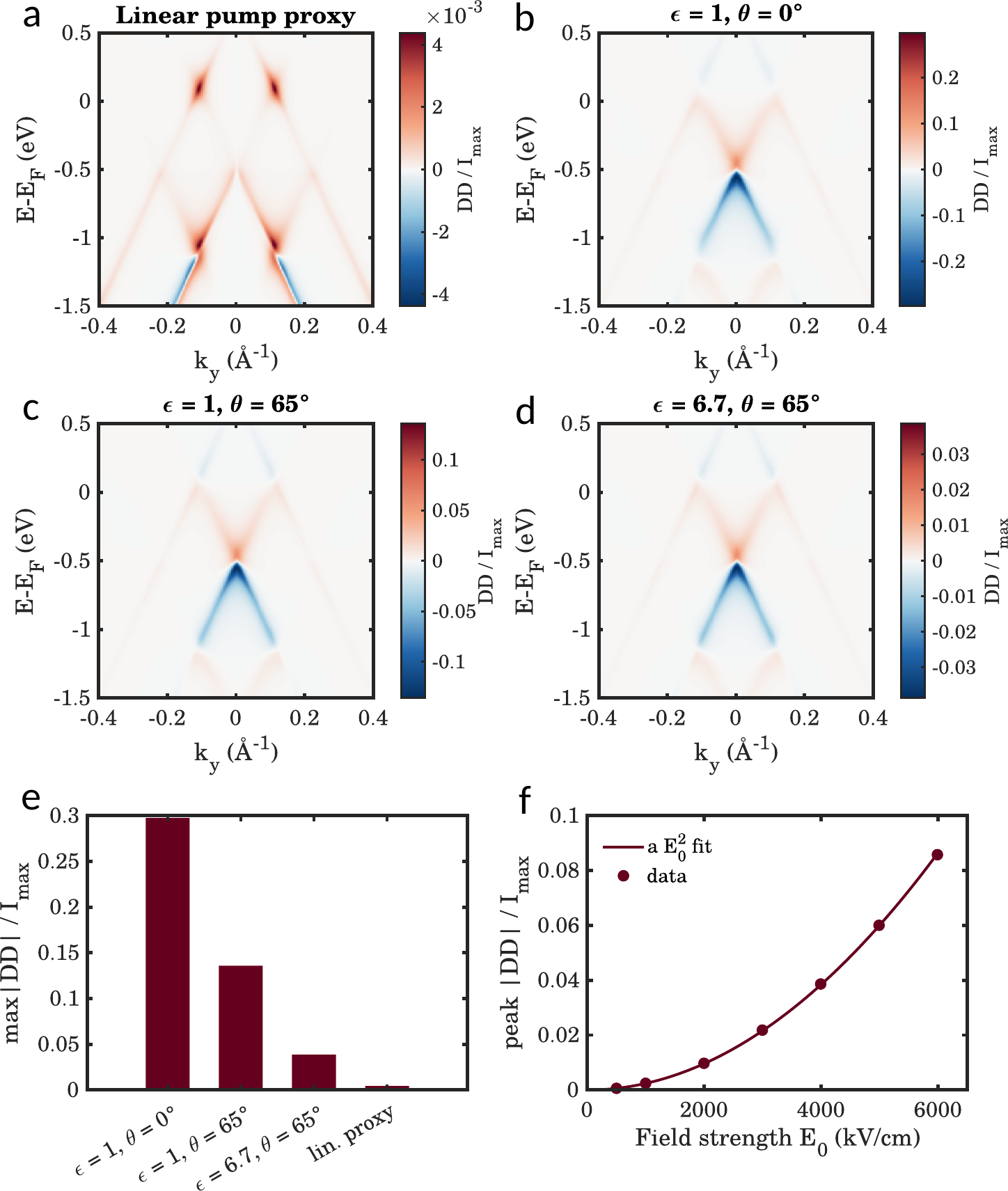}
\caption{\textbf{Topological origin of the DD.} \textbf{a} Simulated DD, using the methodology described in the Methods section, where the effective pump polarization in the $(x,y)$ plane was replaced by its closest linear proxy. \textbf{b--d}: Simulated DD for circular pump polarization (with respect to the vacuum propagation direction) for different dielectric constants $\epsilon$ and angle of incidence $\theta$. \textbf{d} corresponds to the conditions in the experiments. \textbf{e} Quantitative evaluation of the strength of the DD, comparing the different geometry conditions from \textbf{b--d} and the linear proxy case from \textbf{a}. \textbf{f}: Maximum strength of the DD from the full simulation as a function of the peak field strength $E_0$.\label{fig:checks}}
\end{figure}

Because our measurement geometry combines a large angle of incidence with a high-photon-energy XUV probe, we performed extensive sanity checks to confirm that the observed dichroism genuinely originates from light-induced pseudospin texture rather than from a trivial geometric or matrix-element artifact of the oblique, elliptically-polarized drive. First, replacing the physical pump with its closest linear-polarization proxy of identical in-plane field magnitude removes all helicity content, and the resulting proxy dichroism is practically zero (Fig.~\ref{fig:checks}a). 
This confirms that the dichroism is driven by circularity of the pump, not by the incidence geometry itself. For a true circular pump, the details of the geometry and the dielectric constant merely determine the overall scale but not the signature of the topological state in the DD, as demonstrated by the exact same features observed for various conditions in Fig.~\ref{fig:checks}(b--e).

Second, the extracted DD follows the field scaling expected for a second-order (helicity-odd) light–matter process, scaling quadratically with pump field at weak drive (Fig.~\ref{fig:checks}f).

\begin{figure}[H]
\centering
\includegraphics[width=0.95\textwidth]{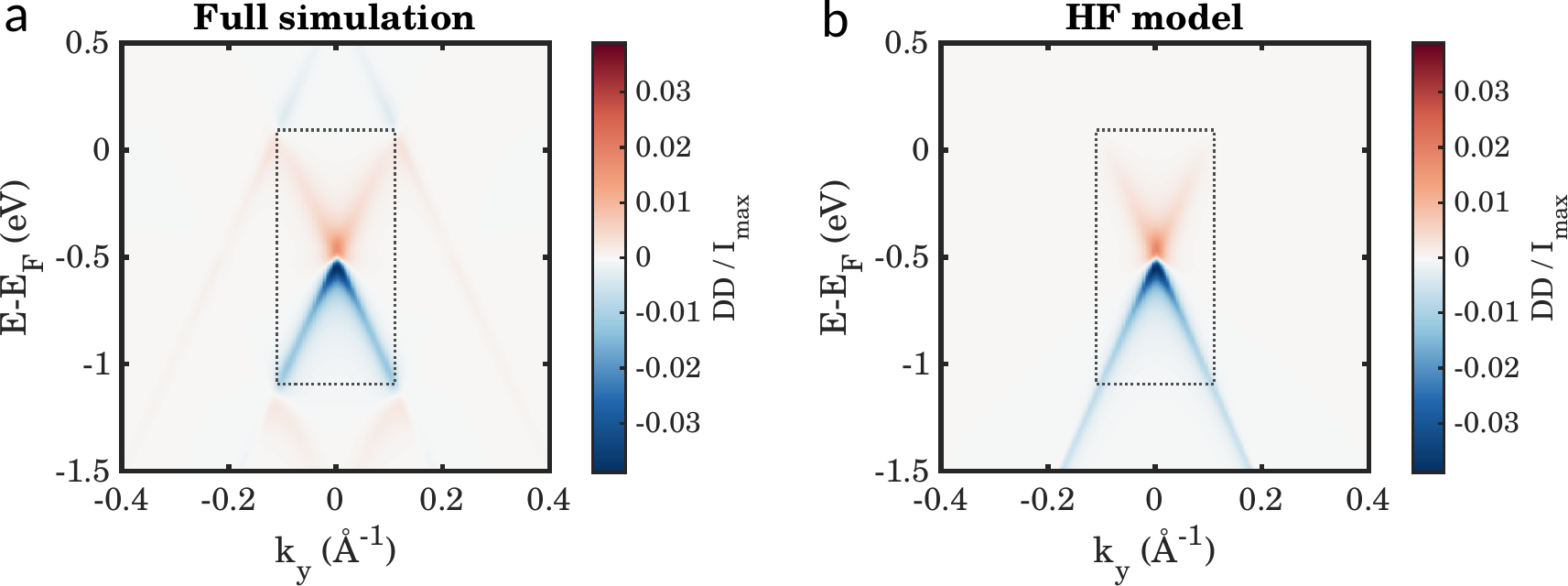}
\caption{\textbf{Pure Floquet origin of the DD.} \textbf{a} DD for the full simulation of the photoemission spectrum from the main text. \textbf{b} DD for the effective Floquet model~\eqref{eq:green_eff} derived from the high-frequency Floquet perturbation theory. We computed the correlation between the full simulation and the effective model for (i) the full energy-momentum region, and (ii) for the region in the vicinity of the Dirac point, indicated by the black dashed box.
\label{fig:floquet_hf}}
\end{figure}

Finally, we isolate the purely helicity-odd, second-order Floquet term responsible for the light-induced Berry curvature -- while discarding all other contributions to the light–matter coupling -- by high-frequency Floquet perturbation theory, Eq.~\eqref{eq:eff_ham} with $\Delta = m_F$. Computing the corresponding effective Green's function
\begin{align}
    \label{eq:green_eff}
    \tilde{G}^\mathrm{R}(\omega) = [\omega - \tilde{H}(\mathbf{k}) + i \eta]^{-1} \ , \
    \tilde{G}^<(\omega) = i f(\omega) [\tilde{G}^\mathrm{R}(\omega) - \tilde{G}^\mathrm{R}(\omega)^\dagger] \ ,
\end{align}
where $f(\omega)$ is the Fermi-Dirac distribution at equilibrium, and inserting into Eq.~(1) in the main text yields the photoemission spectrum for the effective Floquet Hamiltonian. Fig.~\ref{fig:floquet_hf} demonstrates that the DD from the full simulation can be reproduced almost quantitatively by the pure Floquet model, with a correlation of C(full, pure Floquet) = 0.927 in the full considered energy-momentum region, and C(full, pure Floquet) = 0.985 for the region close to the Dirac point (black dashed box in Fig.~\ref{fig:floquet_hf}b).

Taken together, these checks rule out incidental geometric contributions and establish that the observed helicity-reversal dichroism is the direct spectroscopic fingerprint of the light-induced Berry curvature at the Dirac point.

\end{document}